\documentclass[usenatbib]{mn2e}

\usepackage{epsfig}
\usepackage{mathrsfs,aas_macros}
\usepackage{color}

\def\Lya{Ly$\alpha$~}

\def\HI{\hbox{H$\,\rm \scriptstyle I\ $}}

\def\HeII{\hbox{He$\,\rm \scriptstyle II\ $}}

\title[The IGM temperature at redshift $z=2.4$]{A
  consistent determination of the temperature of the intergalactic
  medium at redshift {\boldmath $\langle z \rangle=2.4$}}

\author[J.S. Bolton et al.] {James S. Bolton$^{1}$, George
  D. Becker$^{2}$, Martin G. Haehnelt$^{2}$ \& Matteo
  Viel$^{3,4}$\\$^1$ School of Physics and Astronomy, University of
  Nottingham, University Park, Nottingham, NG7 2RD\\$^2$ Kavli
  Institute for Cosmology and Institute of Astronomy, Madingley Road,
  Cambridge, CB3 0HA\\$^3$ INAF - Osservatorio Astronomico di Trieste,
  Via G.B. Tiepolo 11, I-34131 Trieste, Italy\\$^4$ INFN/National
  Institute for Nuclear Physics, Via Valerio 2, I-34127 Trieste, Italy
}

\begin{document}

\date{\today}

\maketitle

\label{firstpage}

\begin{abstract}

We present new measurements of the thermal state of the intergalactic
medium (IGM) at $z\sim2.4$ derived from absorption line profiles in
the \Lya forest.  We use a large set of high-resolution hydrodynamical
simulations to calibrate the relationship between the
temperature-density ($T$--$\Delta$) relation in the IGM and the
distribution of \HI column densities, $N_{\rm HI}$, and velocity
widths, $b_{\rm HI}$, of discrete \Lya forest absorbers.  This
calibration is then applied to the measurement of the lower cut-off of
the $b_{\rm HI}$--$N_{\rm HI}$ distribution recently presented by
\cite{Rudie12}.  We infer a power-law $T$--$\Delta$ relation,
$T=T_{0}\Delta^{\gamma-1}$, with a temperature at mean density,
$T_{0}=[1.00^{+0.32}_{-0.21}]\times10^{4}\rm\,K$ and slope
$(\gamma-1)=0.54\pm0.11$.  The slope is fully consistent with that
advocated by the analysis of \cite{Rudie12}; however, the temperature
at mean density is lower by almost a factor of two, primarily due to
an adjustment in the relationship between column density and physical
density assumed by these authors.  These new results are in excellent
agreement with the recent temperature measurements of \cite{Becker11},
based on the curvature of the transmitted flux in the \Lya forest.
This suggests that the thermal state of the IGM at this redshift is
reasonably well characterised over the range of densities probed by
these methods.

\end{abstract}



\section{Introduction}
 
 The temperature of the intergalactic medium (IGM) represents a
 fundamental quantity describing the physical state of the majority of
 baryons in the Universe.  In the standard paradigm, the thermal state
 of the low-density IGM, which gives rise to the \Lya forest in quasar
 spectra, is set primarily by the balance between photo-ionisation
 heating and adiabatic cooling.  At redshifts well after reionisation
 completes, this should result in a well-defined power-law
 relationship between temperature and density, $T =
 T_{0}\Delta^{\gamma-1}$, for overdensities $\Delta = \rho/\langle
 \rho \rangle \leq 10$ (\citealt{HuiGnedin97}).  During and
 immediately following hydrogen or helium reionisation, in contrast,
 the temperature-density relation will become multi-valued and
 spatially dependent
 (\citealt{Bolton04,McQuinn09,MeiksinTittley12,Compostella13}).  It
 has been suggested that additional heating processes, such as
 volumetric heating by TeV emission from blazars, may modify this
 picture (\citealt{Puchwein12}, but see \citealt{Miniati13}).
 Regardless of the precise heating mechanism, however, the long
 adiabatic cooling timescale at these low densities, $t_{\rm ad}
 \propto H(z)^{-1}$, means that the IGM retains a long thermal memory.
 The temperature-density ($T$--$\Delta$) relation can therefore serve
 as a powerful diagnostic of the reionisation epoch and the properties
 of ionising sources in the early Universe
 (e.g. \citealt{MiraldaRees94,Theuns02,HuiHaiman03,FurlanettoOh09,Cen09,Raskutti12})

Over the last decade there have been many attempts to measure the
thermal state of the low density IGM using the \Lya forest.  These
measurements may be broadly classed into two approaches: statistical
techniques which treat the \Lya forest transmission as a continuous
field quantity
(\citealt{Zaldarriaga01,Theuns02c,Bolton08,Viel09,Lidz10,Becker11,Garzilli12})
and those that instead decompose the \Lya forest absorption into
discrete line profiles (\citealt{Haehnelt98,
  Schaye00,Ricotti00,BryanMachacek00,
  McDonald01,Bolton10,Bolton12,Rudie12}, hereafter RSP12).  These
studies have produced a variety of results, not all of which are in
good agreement.  The tension between the measurements has historically
remained relatively weak, however, as the statistical and/or
systematic uncertainties have often been large.

Recently, however, two studies based on large data sets have claimed
to measure temperatures in the IGM at $z=2.4$ to high ($<$10 per cent)
precision, but with apparently discrepant results.  The first was by
\cite{Becker11}, who analysed the curvature of the transmitted flux in
the \Lya forest over $2 < z < 5$.  This study measured the temperature
at a characteristic overdensity, $T({\bar \Delta})$, where ${\bar
  \Delta}$ evolved with redshift.  At $z=2.4$, they found $T({\bar
  \Delta})=[2.54\pm0.13] \times 10^{4}\rm\,K$ (2$\sigma$ error) at
${\bar \Delta}=4.4$.  More recently, RSP12 reported a measurement of
the $T$--$\Delta$ relation at $\langle z \rangle=2.37$ using \Lya line
profiles in a large set of high resolution, high signal-to-noise
spectra obtained as part of the Keck Baryonic Structure Survey (KBSS,
\citealt{Rudie12KBSS}).  They measured a temperature at mean density,
$T_{0}=[1.87\pm0.08]\times 10^{4}\rm\,K$, and a power-law slope,
$(\gamma-1)=0.47\pm0.10$ (1$\sigma$ errors) for their ``default''
outlier rejection scheme.  The RSP12 results imply a temperature at
the density probed by \cite{Becker11} of $T({\bar
  \Delta})=[3.75\pm0.58] \times 10^{4}\rm\,K$ (1$\sigma$, including
the errors in both $T_{0}$ and $\gamma-1$), which is discrepant with
the \cite{Becker11} value at over the 2$\sigma$ level.  Stated
differently, for the \cite{Becker11} value of $T({\bar \Delta})$ to be
consistent with the RSP12 value of $T_{0}$ would require $(\gamma-1) =
0.21\pm0.03$, which is at odds with the RSP12 result for the slope of
the $T$--$\Delta$ relation.

Obtaining consistent measurements of the IGM $T$--$\Delta$ relation at
$z=2.4$ would be a considerable step towards establishing the full
thermal history of the high-redshift IGM, which is intimately related
to hydrogen and helium reionisation and to the characteristics of
ionising sources.  For example, the value for the slope of the
$T$--$\Delta$ relation advocated by RSP12 is in good agreement with
that expected for an optically thin, post-reionisation IGM, where the
temperature is set by the balance between adiabatic cooling and
photo-heating only.  This implies there may be no need for the
additional heating processes which have been invoked to explain the
observed probability distribution function (PDF) of the \Lya forest
transmission at $2<z<3$ (\citealt{Bolton08,Viel09}).

One possible avenue towards reconciling the \cite{Becker11} and RSP12
results is to re-examine the calibrations underlying their temperature
results.  \cite{Becker11} employed a suite of hydrodynamical
simulations to convert their curvature measurements to temperatures at
a specific overdensity.  RSP12 obtained their measurement of the
$T$--$\Delta$ relation using the velocity width-column density
($b_{\rm HI}$--$N_{\rm HI}$) cut-off technique developed by
\cite{Schaye99} (hereafter S99).  This latter approach is based on the
premise that the lower envelope of the $b_{\rm HI}$--$N_{\rm HI}$
plane arises from gas which follows the $T$--$\Delta$ relation of the
IGM.  S99 used hydrodynamical simulations to establish and calibrate
the relationship between the $b_{\rm HI}$--$N_{\rm HI}$ cut-off and
the $T$--$\Delta$ relation. In contrast to the original S99 analysis,
however, RSP12 obtained their measurements using an analytical
expression for the relationship between $\Delta$ and $N_{\rm HI}$ and
the assumption that \Lya absorption lines at the $b_{\rm HI}$--$N_{\rm
  HI}$ cut-off are thermally broadened.

In this work we revisit the calibration of the $b_{\rm HI}$--$N_{\rm
  HI}$ cut-off technique used by RSP12.  Specifically, we use a large
set of high resolution hydrodynamical simulations of the IGM to test
the robustness of the values of $T_0$ and $(\gamma-1)$ advocated by
RSP12 based on their measurement of the lower cut-off in the $b_{\rm
  HI}$--$N_{\rm HI}$ distribution.  Our goals are two-fold.  Firstly,
we wish to confront the analytical approach adopted by RSP12 with
detailed simulations of the IGM.  Secondly, using these simulations,
we may assess whether the RSP12 measurement is consistent with other
recent results, or whether there is genuine tension between
measurements of the IGM thermal state at $2<z<3$ obtained using \Lya
transmission statistics and line decomposition techniques.

We note that this redshift range represents an excellent starting
point for establishing a consensus picture of the thermal state of the
high-redshift IGM, for multiple reasons.  First, it is after the end
of helium reionisation (e.g. \citealt{Shull10,Syphers11,Worseck11}),
when the $T$--$\Delta$ relation at low densities can reasonably be
expected to follow a power law.  Second, excellent high-resolution
spectra covering the \Lya forest can be obtained from bright quasars.
Critically, the \Lya forest is also relatively sparse at these
redshifts, enabling a largely straightforward decomposition of the
forest into individual absorption lines and thus an analysis of the
temperature as a function of density using the $b_{\rm HI}$--$N_{\rm
  HI}$ cut-off technique.

We begin by providing a brief overview of the numerical simulations
and methodology in Section~\ref{sec:method}.  Our results are
presented in Section~\ref{sec:results} and a discussion follows in
Section~\ref{sec:discuss}.  A numerical convergence test of the
$b_{\rm HI}$--$N_{\rm HI}$ cut-off measured in the simulations is
presented in an appendix.


\section{Methodology}\label{sec:method}
\subsection{Hydrodynamical simulations and \Lya forest spectra}\label{sec:sims}

In order to calibrate the relationship between the observed $b_{\rm
  HI}$-$N_{\rm HI}$ cut-off and the IGM $T$-$\Delta$ relation, we
utilise cosmological hydrodynamical simulations performed using
\textsc{GADGET-3} (\citealt{Springel05}).  The bulk of the simulations
used in this study are described by \cite{Becker11} (see their table
2). In this work we supplement these models with four additional runs
which explore a finer grid of $(\gamma-1)$ values. This gives a total
of 18 simulations for use in our analysis, all of which have a box
size of $10h^{-1}$ comoving Mpc, a gas particle mass of $9.2\times
10^{4}h^{-1}M_{\odot}$ and follow a wide range of IGM thermal
histories characterised by different $T$-$\Delta$ relations.  The
cosmological parameters adopted in the simulations are $\Omega_{\rm
  m}=0.26$, $\Omega_{\Lambda}=0.74$, $\Omega_{\rm b}h^{2}=0.023$,
$h=0.72$, $\sigma_{8}=0.80$, $n_{\rm s}=0.96$, with a helium mass
fraction of $Y=0.24$.  We shall demonstrate later our results remain
insensitive to this choice by instead using the recent Planck results
(\citealt{planck2013}).

Mock \Lya forest spectra are extracted from the simulations from
outputs at $z=2.355$ and are processed to match the RSP12
observational data (e.g. \citealt{Theuns98}).  The mean transmitted
flux, $\langle F \rangle$, of the spectra is matched to the recent
measurements presented by \cite{Becker13}.  The simulated data are
then convolved with a Gaussian instrument profile with $\rm
FWHM=7\rm\,km\,s^{-1}$ and rebinned onto pixels of width $\sim
3\rm\,km\,s^{-1}$ to match the Keck/HIRES spectrum characteristics of
RSP12.  Finally, Gaussian distributed noise is added with a total
signal-to-noise of $\rm S/N=89$ pixel, matching the mean $\rm S/N$ of
the RSP12 data.

\subsection{From $b_{\rm HI}$--$N_{\rm HI}$ cut-off to $T$--$\Delta$ relation}

The procedure used for measuring the $b_{\rm HI}$--$N_{\rm HI}$
cut-off in the simulations follows the method described in S99 and
RSP12.  We use \textsc{VPFIT}\footnote{Version 10.0 by R.F. Carswell
  and J.K. Webb, http://www.ast.cam.ac.uk/$\sim$rfc/vpfit.html} to fit
Voigt profiles to our mock \Lya forest spectra.  We then select a
sub-sample of the lines using the ``default'' RSP12 selection
criteria: only absorbers with $8\leq b_{\rm HI}/[{\rm km\,s^{-1}}]\leq
100$, $10^{12.5} \leq N_{\rm HI}/[{\rm cm^{-2}}]\leq10^{14.5}$ and
relative errors of less than 50 per cent are included.  We furthermore
ignore all lines within $100\,\rm\,km\,s^{-1}$ of the edges of our
mock spectra to avoid possible line duplications arising from the
periodic nature of the simulation boundaries.  We note, however, that
we opted not to use the additional $\sigma$-rejection scheme
introduced by RSP12.  We found this procedure did not work effectively
when applied to models with $b_{\rm HI}$--$N_{\rm HI}$ distributions
which differed significantly from the RSP12 data.  We therefore always
compare to the RSP12 ``default'' measurements throughout this work.

The power-law cut-off at the lower envelope of the $b_{\rm
  HI}$--$N_{\rm HI}$ plane, $b_{\rm HI}=b_{\rm HI,0}(N_{\rm HI}/N_{\rm
  HI,0})^{\Gamma-1}$, is then measured in each of our simulations by
applying the iterative fitting procedure described by S99 to $4\,000$
\Lya absorption lines selected following the above criteria (the full
RSP12 sample contains $5\,758$ \HI absorbers).  Uncertainties on
$b_{0}$ and $(\Gamma-1)$ are estimated by bootstrap sampling the lines
with replacement $4\,000$ times.

Once  the $b_{\rm HI}$--$N_{\rm HI}$ cut-off has been measured in
the simulations, the $T$--$\Delta$ relation may then be inferred by
utilising the ansatz tested by S99 -- namely that power-law
relationships hold between $\Delta$--$N_{\rm HI}$ and $T$--$b_{\rm
  HI}$ for absorption lines near the cut-off, such that,

\begin{equation}  \log \Delta = \zeta_{1} + \xi_{1}\log(N_{\rm HI}/N_{\rm HI,0}), \label{eq:delta} \end{equation}
\begin{equation}  \log T = \zeta_{2} + \xi_{2}\log b_{\rm HI}.  \label{eq:temp}\end{equation}

\noindent
Substituting these expressions into the $T$--$\Delta$ relation and
comparing coefficients with the power-law cut-off, $b_{\rm HI}=b_{\rm
  HI,0}(N_{\rm HI}/N_{\rm HI,0})^{\Gamma-1}$, we identify, 

\begin{equation} \gamma - 1 = \frac{\xi_{2}}{\xi_{1}}(\Gamma-1), \label{eq:r1} \end{equation}
\begin{equation} \log T_{0}= \xi_{2}\log b_{\rm HI,0} + \zeta_{2} - \zeta_{1}(\gamma-1), \label{eq:r2} \end{equation}

\noindent
where $\zeta_{1}=0$ when $N_{\rm HI,0}$ is the column density
corresponding to gas at mean density.  A measurement of $T_{0}$ and
$(\gamma-1)$ therefore relies on correctly identifying the remaining
three coefficients in Eqs.~(\ref{eq:delta}) and~(\ref{eq:temp}), and
\emph{additionally}, $N_{\rm HI,0}$.  Selecting a value of $N_{\rm
  HI,0}$ which is too high or low will systematically bias the
inferred $T_{0}$ unless the $T$--$\Delta$ relation is isothermal
(i.e. $\gamma-1=0$).

The analysis presented by RSP12 assumed $b_{\rm HI}=(2k_{\rm
  B}T/m_{\rm H})^{1/2}$ (i.e. that lines near the cut-off are thermally
broadened only) and\footnote{Eq.~(\ref{eq:schaye01}) ignores the weak
  dependence of the column density on the slope of the $T$--$\Delta$
  relation, $N_{\rm HI}\propto \Delta^{3/2 -
    0.22(\gamma-1)}T_{0,4}^{-0.22}$.  Note also the normalisation of
  this expression is $0.07$ dex lower than eq. (2) in RSP12.  This
  difference is due to the slightly different cosmological parameters
  and case-A recombination coefficient, $\alpha_{\rm HII}=4.063\times
  10^{-13}T_{4}^{-0.72}\rm\,cm^{3}\,s^{-1}$, assumed in this work.}

\begin{equation} N_{\rm HI} \simeq 10^{13.23}\rm\,cm^{-2}\Delta^{3/2} \frac{T_{4}^{-0.22}}{\Gamma_{-12}}\left( \frac{1+z}{3.4}\right)^{9/2}, \label{eq:schaye01} \end{equation}

\noindent
where $T_{4}=T/10^{4}\rm\,K$ and $\Gamma_{-12}=\Gamma_{\rm
  HI}/10^{-12}\rm\,s^{-1}$ is the metagalactic \HI photo-ionisation
rate.  Eq.~(\ref{eq:schaye01}) assumes the typical size of a \Lya
forest absorber is the Jeans scale (\citealt{Schaye01}) and that the
low-density IGM is in photo-ionisation equilibrium.  With these
assumptions, $\zeta_{1}=0$, $\zeta_{2}=1.782$, $\xi_{1}=2/3$ and
$\xi_{2}=2$ (assuming $b_{\rm HI}$ is in units of $\rm
km\,s^{-1}$). RSP12 furthermore assumed $N_{\rm HI,0}=10^{13.6}\rm\,cm^{-2}$,
based on their evaluation of Eq.~(\ref{eq:schaye01}) for $T_{4}=1$ and
$\Gamma_{-12}=0.5$.


\section{Results} \label{sec:results}

\subsection{The $b_{\rm HI}$--$N_{\rm HI}$ cut-off measured from hydrodynamical simulations}

\begin{figure*}
  \centering
  \begin{minipage}{180mm}
    \begin{center}
      \psfig{figure=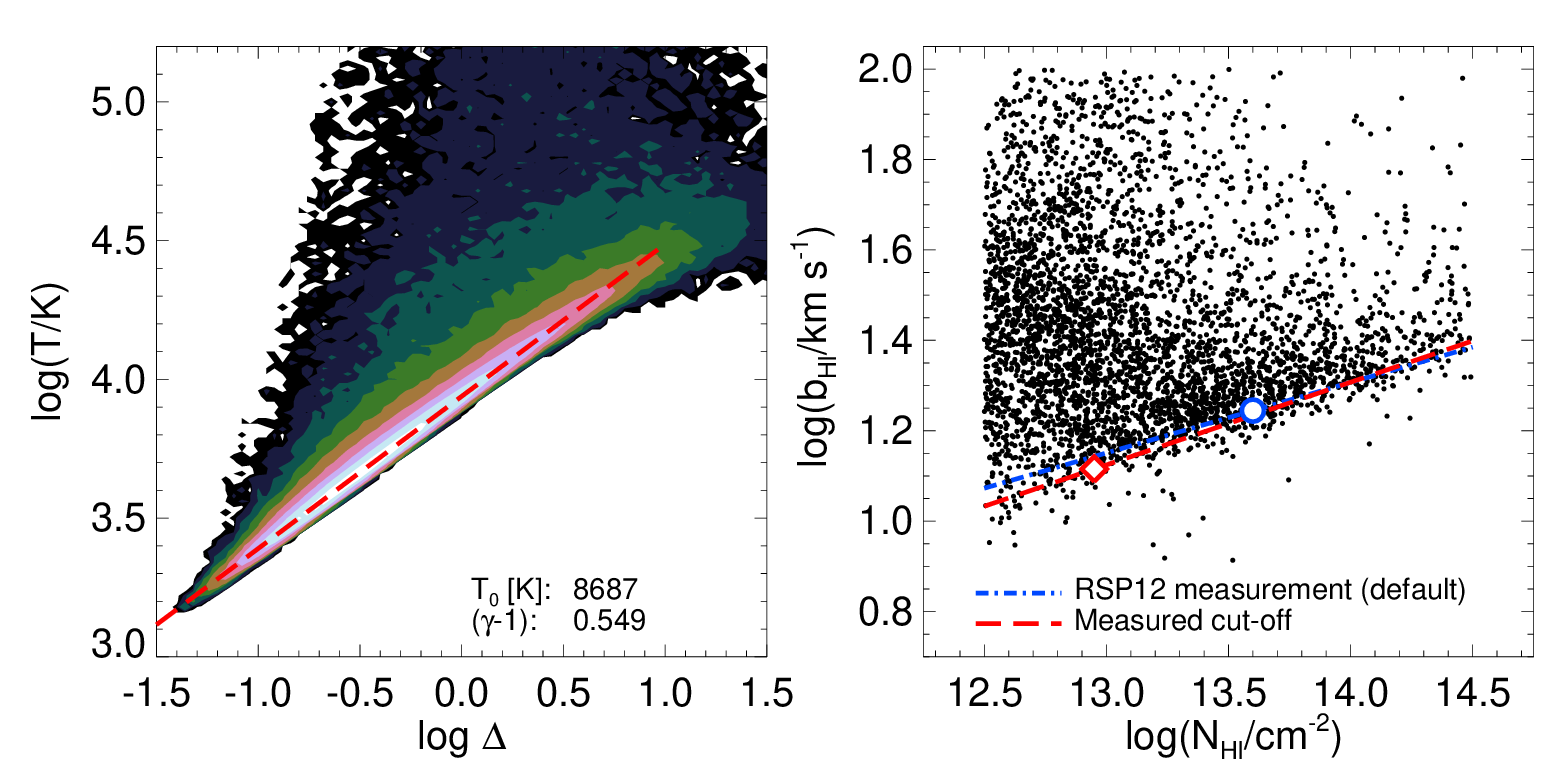,width=0.65\textwidth}
      \psfig{figure=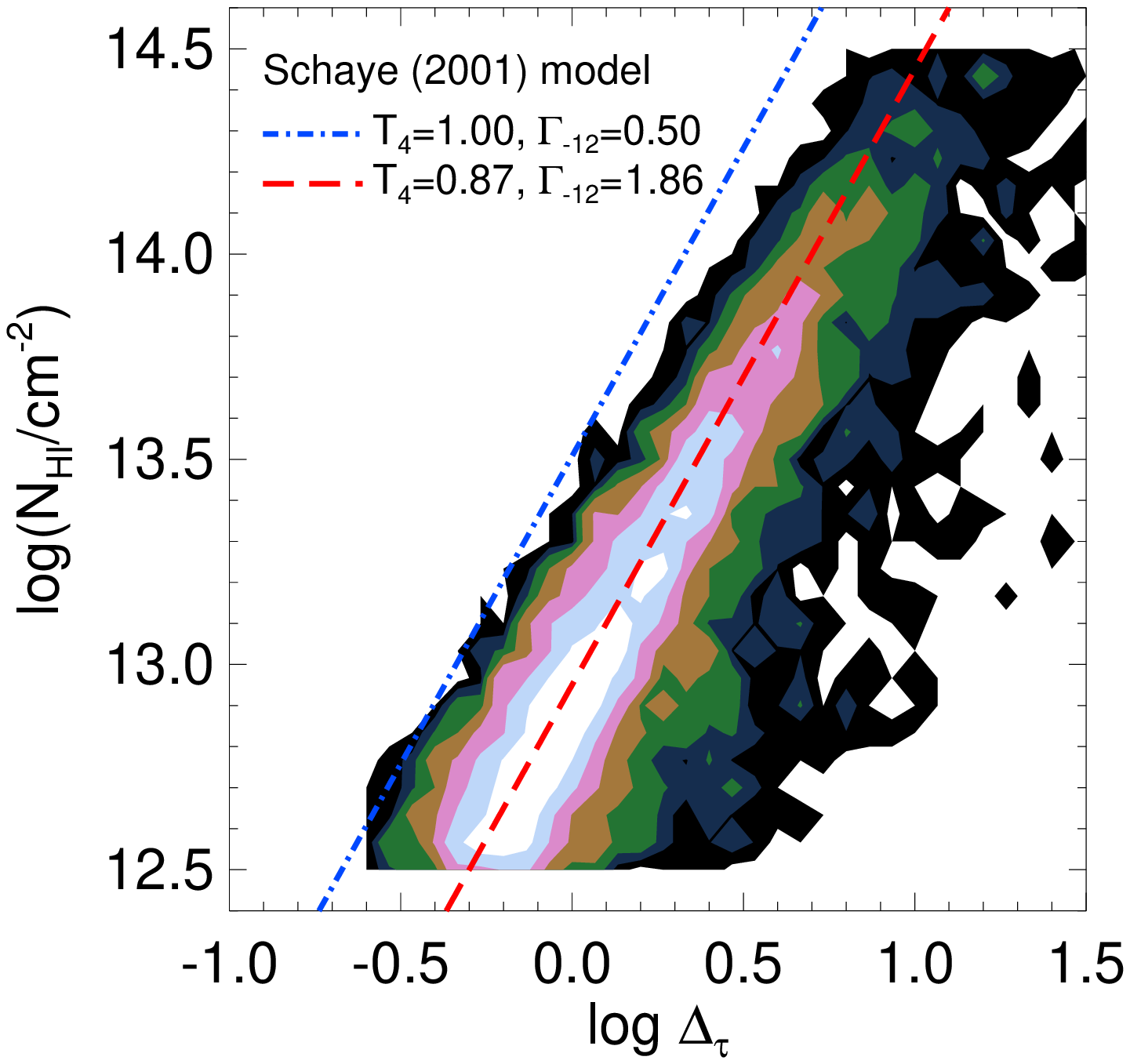,width=0.33\textwidth}
      \psfig{figure=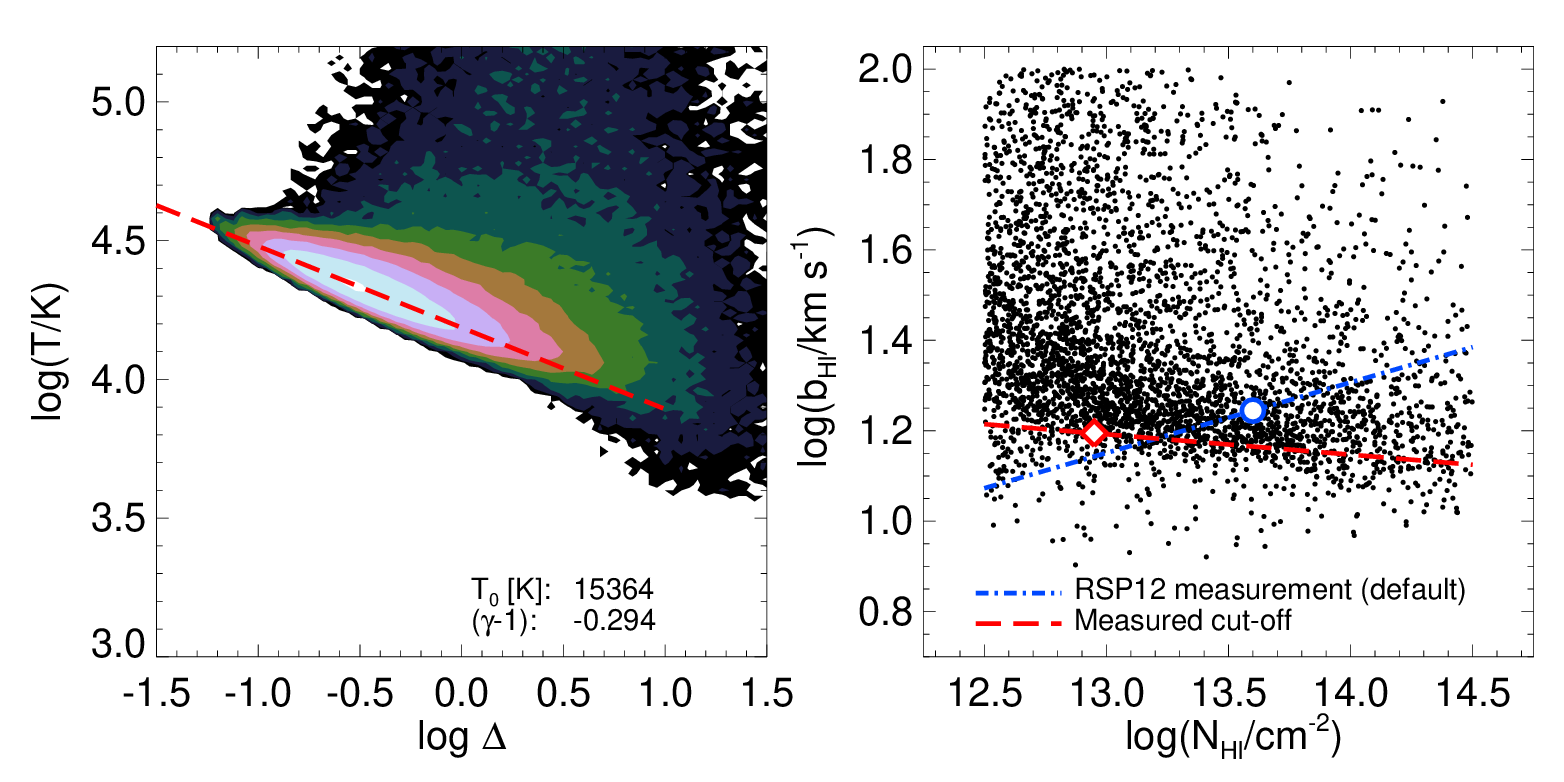,width=0.65\textwidth}
      \psfig{figure=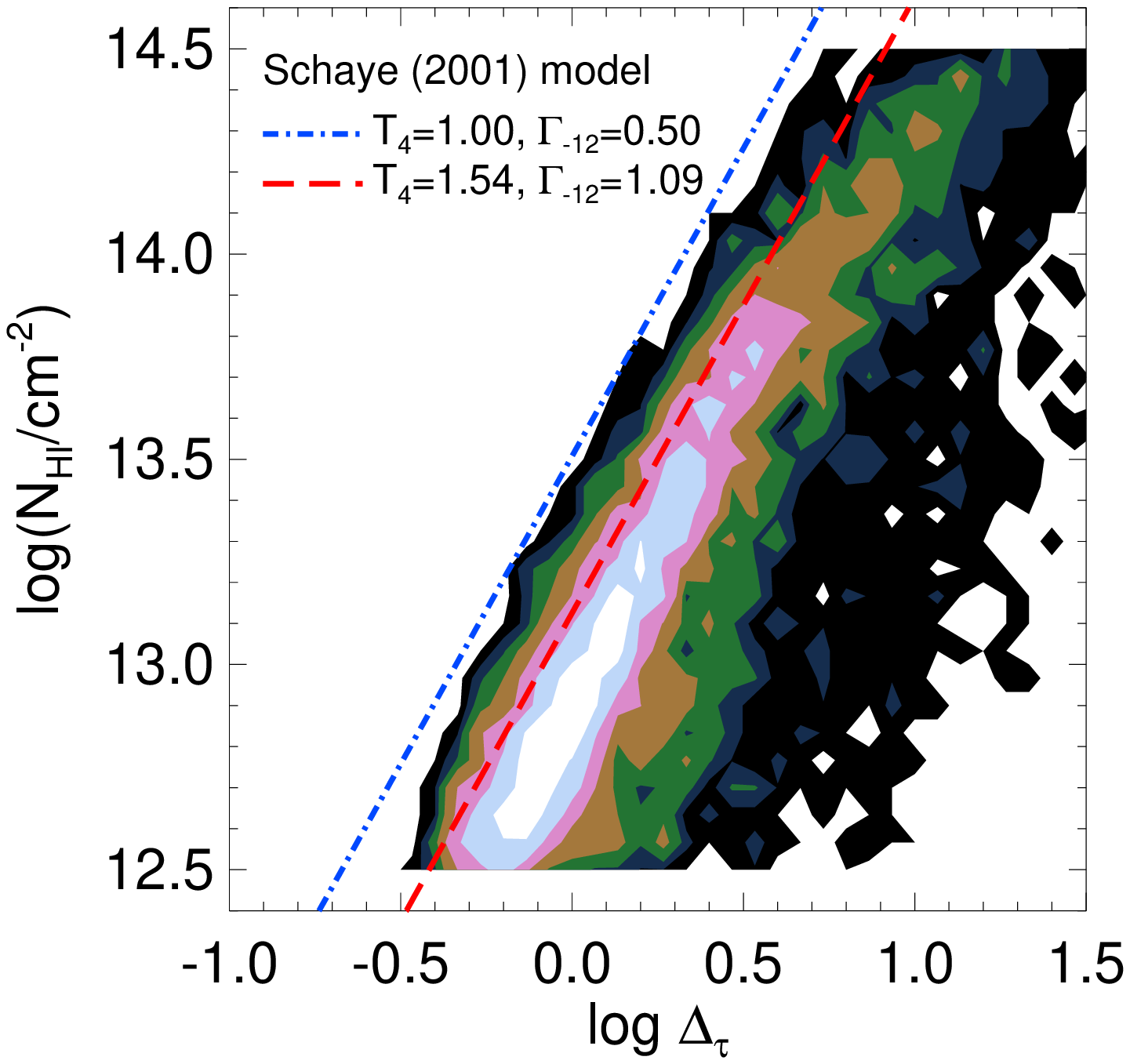,width=0.33\textwidth}
      
      \vspace{-0.3cm}
      \caption{{\it Top left:} Contour plot of the volume weighted IGM
        $T$--$\Delta$ plane in one of the hydrodynamical simulations
        used in this study. The number density of the data points
        increases by $0.5$ dex with each contour level.  The red
        dashed line displays the power-law approximation,
        $T=T_{0}\Delta^{\gamma-1}$, to the $T$--$\Delta$ relation at
        $\log \Delta \leq 1$. {\it Top middle:} The corresponding
        $b_{\rm HI}$--$N_{\rm HI}$ plane for $4\,000$ absorption
        features identified in mock \Lya forest spectra drawn from the
        simulation. The red dashed line corresponds to the measured
        $b_{\rm HI}$--$N_{\rm HI}$ cut-off, while the blue dot-dashed
        line shows the observational measurement presented by
        RSP12. The red diamond indicates the column density
        corresponding to gas with an optical depth weighted
        overdensity of $\log \Delta_{\tau}=0$ in the simulation,
        $N_{\rm HI,0}$, and the blue circle shows the value assumed by
        RSP12.  {\it Top right:} Contour plot of $N_{\rm HI}$ against
        the optical depth weighted gas overdensity at the line
        centres.  The number density of data points increases by
        $0.25$ dex within each contour level.  The red dashed curve
        displays the analytical model of \citet{Schaye01}, evaluated
        using the parameters adopted in the simulation.  The blue
        dot-dashed line instead assumes $T_{4}=1$ and
        $\Gamma_{-12}=0.5$. {\it Bottom:} As for the upper panels, but
        now for a simulation with an inverted $T$--$\Delta$ relation.
        Note that the blue dot-dashed line in the middle panel again
        shows the observational measurement from RSP12.}

      \label{fig:bNfits}
    \end{center}
  \end{minipage}
\end{figure*}

Measurements of the $b_{\rm HI}$--$N_{\rm HI}$ cut-off obtained from
two of our hydrodynamical simulations are displayed in
Fig.~\ref{fig:bNfits}.  The left-most panels display the $T$--$\Delta$
relation in the simulations.  The power-law relationship between the
gas at the lower boundary of the $T$--$\Delta$ plane, shown by the red
dashed lines, is estimated by finding the mode of the gas temperature
in density bins of width 0.02 dex at $\log \Delta = 0$ and $-0.75$.
The middle panels show the corresponding $b_{\rm HI}$--$N_{\rm HI}$
plane and the measured cut-off (red dashed lines).  For comparison,
the RSP12 observational measurement is displayed by the blue
dot-dashed lines; {\it note this is shown in both of the middle panels
  and is not a fit to the simulation data.}  The simulation in the top
panels has a maximally steep $T$--$\Delta$ relation with
$(\gamma-1)=0.55$, while the bottom panels display a model with an
inverted $T$--$\Delta$ relation, $(\gamma-1)=-0.29$.  This qualitative
comparison indicates that a $T$--$\Delta$ relation which is inverted
over $10^{12.5}\leq N_{\rm HI}/[\rm\,cm^{-2}] \leq 10^{14.5}$ is
indeed inconsistent with the Voigt profile fits to the KBSS data, in
agreement with the conclusion of RSP12.  However, the value of $N_{\rm
  HI,0}$ in the simulations, indicated by the red diamond in the
middle panels, is $\sim 0.65$ dex smaller than the value used by
RSP12, shown by the blue circle.

The explanation for this becomes apparent on examining the right-most
panels in Fig.~\ref{fig:bNfits}, which display the relationship
between $N_{\rm HI}$ and the corresponding optical depth-weighted
overdensities at the line centres.\footnote{Following S99, we match
  the column densities measured with VPFIT to the optical depth
  weighted overdensity of the gas at the line centre,
  $\Delta_{\tau}=\sum\tau_{\rm i}\Delta_{\rm i}/\sum\tau_{\rm i}$, where the
  summation is over all pixels, $i$, along a simulated sight-line.
  This mitigates the effect of redshift space distortions arising from
  peculiar motions and line broadening, which would otherwise distort
  the direct mapping between $N_{\rm HI}$ and $\Delta$ in real space.}
The blue dot-dashed lines display Eq.~(\ref{eq:schaye01}) assuming
$T_{0,4}=1$ and $\Gamma_{-12}=0.5$.  The slope of this relation is in
excellent agreement with the simulation data, implying that a power
law relationship between $\Delta$ and $N_{\rm HI}$, as inferred by
\cite{Schaye01}, is a very good approximation.

On the other hand, the normalisation of this relation disagrees with
the simulations.  Since the temperature dependence of
Eq.~(\ref{eq:schaye01}) is weak, this difference is mainly due to the
low value of $\Gamma_{-12}=0.5$ RSP12 used to estimate $N_{\rm HI,0}$.
This was based on the results of \cite{Faucher08b}, who inferred
$\Gamma_{-12}$ from the \Lya forest opacity assuming an IGM
temperature at mean density of $T_{0}=2.1\times 10^{4}\rm\,K$
(\citealt{Zaldarriaga01}).  As \cite{Faucher08b} correctly point out,
the \Lya forest opacity constrains the quantity
$T_{0}^{-0.72}/\Gamma_{-12}$, and so assuming a larger (smaller) IGM
temperature\footnote{The IGM temperature assumed by \cite{Faucher08b}
  is based on the results of \cite{Zaldarriaga01}, who inferred
  $T_{0}$ from the \Lya forest power spectrum after calibrating their
  measurement with a dark matter only simulation performed with a
  (now) outdated cosmology.  Since $\tau_{\rm Ly\alpha}\propto
  T_{0}^{-0.72}/\Gamma_{-12}$, it is therefore not entirely
  coincidental that the RSP12 constraint on $T_{0}$ is similar to the
  measurement presented by \cite{Zaldarriaga01}.}  at mean density
will translate their constraint into a smaller (larger) value of
$\Gamma_{-12}$ if the \Lya opacity remains fixed.  In addition,
\cite{Faucher08b} obtained their $\Gamma_{-12}$ measurements using an
analytical model for IGM absorption which ignored the effect of
redshift space distortions on the \Lya forest opacity.  Including
peculiar velocities and line broadening, as we do with our
simulations, raises the $\Gamma_{-12}$ required to match a given value
of the mean transmission in the \Lya forest by up to 30 per cent
(\citealt{BeckerBolton13}).  The \cite{Faucher08b} measurements are
therefore systematically lower (by a factor of two or more) compared
to the value we require to match our mock spectra to the mean
transmission measurements of \cite{Becker13} if $T_{0}\sim 1\times
10^{4}\rm\,K$.

This is demonstrated by the red dashed lines in the right-hand panels
of Fig.~\ref{fig:bNfits}, which display Eq.~(\ref{eq:schaye01})
evaluated using the photo-ionisation rates and gas temperatures used
in the simulations. Although the agreement between the analytical
model and simulations is still not perfect\footnote{Note that $\Delta$
  and $\Delta_{\tau}$ are slightly different quantities, and this
  comparison is thus not exact.}, the higher $\Gamma_{-12}$ values
result in a significantly improved correspondence.  We therefore
conclude that the value $N_{\rm HI,0}=10^{13.6}\rm\,cm^{-2}$ assumed
by RSP12 is biased high by $\sim 0.65$ dex.  As we will demonstrate,
this bias translates into a significant overestimate of $T_0$.
Finally, note that the 68 (95) per cent bounds on the range of optical
depth weighted overdensities probed by absorption lines with column
densities $10^{12.5}\leq N_{\rm HI}/[\rm\,cm^{-2}] \leq 10^{14.5}$ are
$-0.2\la \log \Delta_{\tau} \la 0.7$ ($-0.4\la \log \Delta_{\tau} \la
1.2$) at $z=2.4$. The $b_{\rm HI}$--$N_{\rm HI}$ cut-off approach will
be largely insensitive to the slope of the $T$--$\Delta$ relation
outside this range of overdensities.

\subsection{A re-evaluation of the inferred $T$--$\Delta$ relation at $\langle z\rangle =2.37$}

\begin{figure}
\begin{center}
  \includegraphics[width=0.495\textwidth]{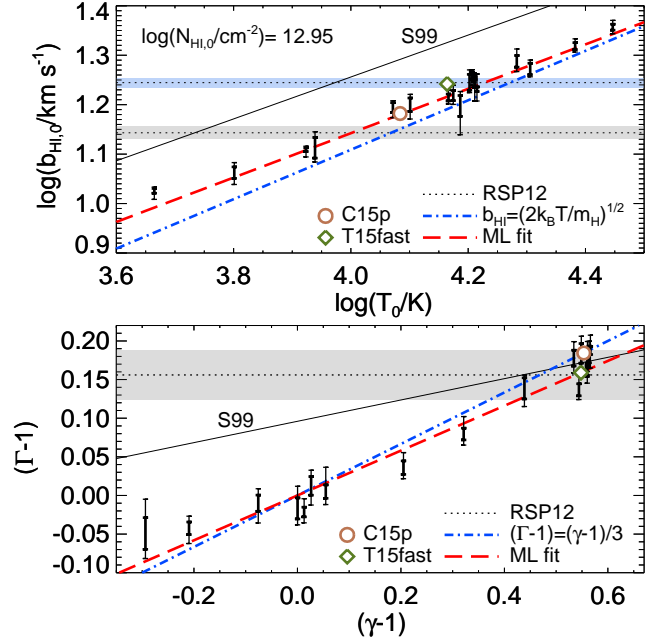}
  \vspace{-0.5cm}
  \caption{Measurements of the amplitude $b_{\rm HI,0}$ (upper panel)
    and slope $(\Gamma-1)$ (lower panel) of the $b_{\rm HI}$--$N_{\rm
      HI}$ cut-off against the corresponding $T_{0}$ and $(\gamma-1)$
    from $18$ hydrodynamical simulations.  The thick (thin) bootstrap
    error bars correspond to the 68 (95) per cent confidence intervals
    around the median.  The red dashed lines give the maximum
    likelihood power-law fits to the data using the bootstrapped
    uncertainty distributions, while the blue dot-dashed lines display
    the analytical relations used by RSP12. The solid black lines show
    the power-law relations S99 inferred from their hydrodynamical
    simulations (note these are obtained at $z=3$ rather than
    $z=2.4$), and the grey shaded bands display the RSP12 default
    measurements.  Note that the RSP12 measurement of $b_{\rm HI,0}$
    has been rescaled to correspond to $N_{\rm
      HI,0}=10^{12.95}\rm\,cm^{-2}$; the light blue band gives the
    value measured by RSP12 when assuming $N_{\rm
      HI,0}=10^{13.6}\rm\,cm^{-2}$.  The models C$15_{\rm P}$ and
    T15fast refer to two additional simulations which test the effect
    of cosmological parameters and Jeans smoothing (see text for
    details).  These latter two models were not included in the
    maximum likelihood fits. }

  \label{fig:results}
\end{center}
\end{figure}

We now turn to the key result of this work, summarised in
Fig.~\ref{fig:results}, which displays the relationship between
$b_{\rm HI,0}$--$T_{0}$ (upper panel) and $(\gamma-1)$--$(\Gamma-1)$
(lower panel) obtained from the hydrodynamical simulations.  Our
choice of $N_{\rm HI,0}=10^{12.95}\rm\,cm^{-2}$ corresponds to the
average value of $N_{\rm HI}$ associated with gas with $\log
\Delta_{\tau}=0$ in all 18 simulations used in the analysis.  In
practice, $N_{\rm HI,0}$ varies slightly from one simulation to the
next due to the weak dependence of the column density on the thermal
state of the gas, ranging from $10^{12.8}\rm\,cm^{-2}$ to
$10^{13.1}\rm\,cm^{-2}$ in our coldest to hottest simulations.  We
found adopting values of $N_{\rm HI,0}$ outside of this range
introduces an increasingly significant scatter into the $b_{\rm
  HI,0}$--$T_{0}$ correlation (see also fig. 5 in \citealt{Schaye00}),
invalidating the assumption that Eq.~(\ref{eq:r2}) is a single
power-law (i.e. that $\zeta_{1}=0$ holds) and biasing any measurement
of $T_{0}$ if $(\gamma-1) \neq 0$.

The dashed red lines in Fig.~\ref{fig:results} display the best fit
power-laws to the simulation data given by Eqs.~(\ref{eq:delta}) and
(\ref{eq:temp}), where $\zeta_{2}=1.46$, $\xi_{1}=0.65$ and
$\xi_{2}=2.23$ assuming $\zeta_{1}=0$ and $N_{\rm
  HI,0}=10^{12.95}\rm\,cm^{-2}$.  The blue dot-dashed lines display
the analytical relations used by RSP12.  The dotted lines with shaded
error regions show the RSP12 measurements for two value of $N_{\rm
  HI,0}$.  The grey bands correspond to the case where we have
rescaled the default RSP12 $b_{\rm HI,0}$ to the value measured at
$N_{\rm HI}=10^{12.95}\rm\,cm^{-2}$.  The light blue band in the upper
panel of Fig.~\ref{fig:results} shows their original measurement
assuming $N_{\rm HI,0}=10^{13.6}\rm\,cm^{-2}$.  The best-fit $b_{\rm
  HI,0}$--$T_{\rm 0}$ relation lies slightly above the result for pure
thermal broadening, indicating that additional processes such as
  Jeans smoothing (which also scales as $T^{1/2}$, see
  e.g. \citealt{GnedinHui98}) and Hubble broadening impact on the
  minimum line width.

For comparison, the solid black lines display the relationship at
$z=3$ found from the simulations performed by S99. These authors
inferred a similar slope for the $b_{\rm HI,0}$--$T_{\rm 0}$ relation,
but with a positive offset of around $\sim 0.1\rm\,dex$ relative to
this work.  Aside from the slightly lower redshift we consider here,
there are several possible explanations for this result.  The first is
the smaller dynamic range of the hydrodynamical simulations used by
S99, which employed a gas particle mass of $1.65\times
10^{6}(\Omega_{\rm b}h^{2}/0.0125)(h/0.5)^{-3}M_{\odot}$ within a box
size of $2.5h^{-1}\rm\,Mpc$ and were performed with a modified version
of the smoothed particle hydrodynamics code \textsc{HYDRA}
(\citealt{Couchman95}).  Additionally, the six simulations S99
analysed all used different assumptions for cosmological parameters
and the IGM thermal history, which further complicates a direct
comparison.  A final possibility is that there are differences in the
version of \textsc{VPFIT} S99 used to fit the absorption lines.  The
results presented by S99 already indicated that the assumption of
purely thermal broadening may result in an overestimate of the gas
temperature, but our analysis suggests that the factor by which the
velocity widths of the lines at the $b_{\rm HI}$--$N_{\rm HI}$ cut-off
are actually increased by non-thermal broadening is rather small.

The relatively good agreement of the slopes of the best-fit and
analytical relations in Fig.~\ref{fig:results} reflects the fact that
both Eq.~(\ref{eq:schaye01}) and the assumption that $b_{\rm
  HI}\propto T^{1/2}$ at the $b_{\rm HI}$--$N_{\rm HI}$ cut-off
capture the results of simulations quite accurately.  Interestingly,
however, although the slope of the $b_{\rm HI}$--$T_{0}$ relation is
similar to that found by S99 at $z=3$, we find a much steeper slope
for the relationship between $(\Gamma-1)$ and $(\gamma-1)$. S99
concluded the weaker dependence they found between $(\Gamma-1)$ and
$(\gamma-1)$ would hamper any precise measurement of the slope of the
$T$--$\Delta$ relation.  Our results instead suggest that at $z\sim
2.4$ the slope of the $b_{\rm HI}$--$N_{\rm HI}$ cut-off is able to
discriminate rather well between differing $T$--$\Delta$ relations.
We have checked that using the smaller line sample (300 lines, with
500 bootstrap resamples) used by S99 increases the uncertainties on
the cut-off measurements but does not change this conclusion.  The
exact explanation for this difference is again unclear, although we
again speculate differences in the hydrodynamical simulations may play
a role.  However, the potentially greater sensitivity to $(\gamma-1)$
provides additional motivation for revisiting $b_{\rm HI}$--$N_{\rm
  HI}$ cut-off measurements at higher redshifts.

We have also verified that the best-fit power-law relations in
Fig.~\ref{fig:results} should be robust to small differences in the
assumed cosmology and uncertainties in the pressure (Jeans) smoothing
scale of gas in the IGM (see \citealt{Rorai13} for a recent discussion).
The circles show $b_{\rm HI,0}$ and $(\Gamma-1)$ measured from an
additional simulation, C15$_{\rm P}$ (see \citealt{BeckerBolton13}),
which was performed with cosmological parameters consistent with the
Planck results, $\Omega_{\rm m}=0.308$, $\Omega_{\Lambda}=0.692$,
$\Omega_{\rm b}h^{2}=0.0222$, $h=0.678$, $\sigma_{8}=0.829$ and
$n_{\rm s}=0.961$ (\citealt{planck2013}).  The diamonds show the
results obtained from the simulation T15fast described in
\cite{Becker11}.  This simulation rapidly heats the IGM from
$z=3.5$--$3.0$ by $\sim 9\,000\rm \,K$, in contrast to the more
gradual heating used in our fiducial simulations.  The good agreement
of both these models with the best-fit relations indicates that
neither of these issues should be a significant concern in our
analysis.

Combining these results with Eqs.~(\ref{eq:r1})--(\ref{eq:r2}) and
applying them to the default RSP12 measurement of $b_{\rm HI,0}$ and
$(\Gamma-1)$, we infer $T_{0}= [1.00^{+0.32}_{-0.21}]\times 10^{4}
\rm\,K$ and $(\gamma-1)=0.54\pm 0.11$ at $\langle z \rangle = 2.37$
(cf.  $T_{0}= [1.87\pm 0.08]\times 10^{4} \rm\,K$ and
$\gamma-1=0.47\pm 0.10$ from RSP12).  Note that \emph{we do not
  recompute the RSP12 statistical uncertainty estimates}; we instead
simply rescale their published measurements of $b_{\rm HI,0}$ and
$(\Gamma-1)$ using the results of our simulations.  We have, however,
added (in quadrature) an additional systematic error to our
measurement of $T_{0}$ by assuming an uncertainty of $\pm0.2$ dex in
$\log \Delta$, the fractional overdensity corresponding to the column
density $N_{\rm HI,0}$ at which $T_{0}$ is measured (i.e. $\log
\Delta=0.0\pm0.2$).  This accounts for the intrinsic scatter in the
relationship between $N_{\rm HI}$ and $\Delta_{\tau}$ in our
simulations (see Fig.~\ref{fig:bNfits}) as well as the small
uncertainty in the mean transmitted flux in the \Lya forest
(\citealt{Becker13}).

Our recalibrated measurement for the slope of the $T$--$\Delta$
relation is fully consistent with RSP12.  Importantly, however, we
find their measurement of $T_{0}$ is biased high by around
$9\,000\rm\,K$. This is primarily due the value of $N_{\rm
  HI,0}=10^{13.6}\rm\,cm^{-2}$ they adopt in their analysis -- this
column density corresponds to gas with $\Delta\sim2$--$4$ in our
simulations -- and to a smaller extent their assumption of pure
thermal broadening at the $b_{\rm HI}$--$N_{\rm HI}$ cut-off.
Finally, we stress we did not reanalyse the RSP12 observational data
in this work.  As a result we are unable to fully assess the
importance of other potential systematics such as metal line
contamination, spurious line fits or bias due to differences in the
line fitting procedure (e.g. \citealt{Rauch93}).  Ideally, potential
biases arsing from the latter two possibilities should be minimised by
applying exactly the same line profile fitting procedure to both the
observations \emph{and} simulations (e.g. S99, \citealt{Bolton12}).

\subsection{Comparison to previous measurements} \label{sec:previous}

\begin{figure}
\begin{center}
  \includegraphics[width=0.49\textwidth]{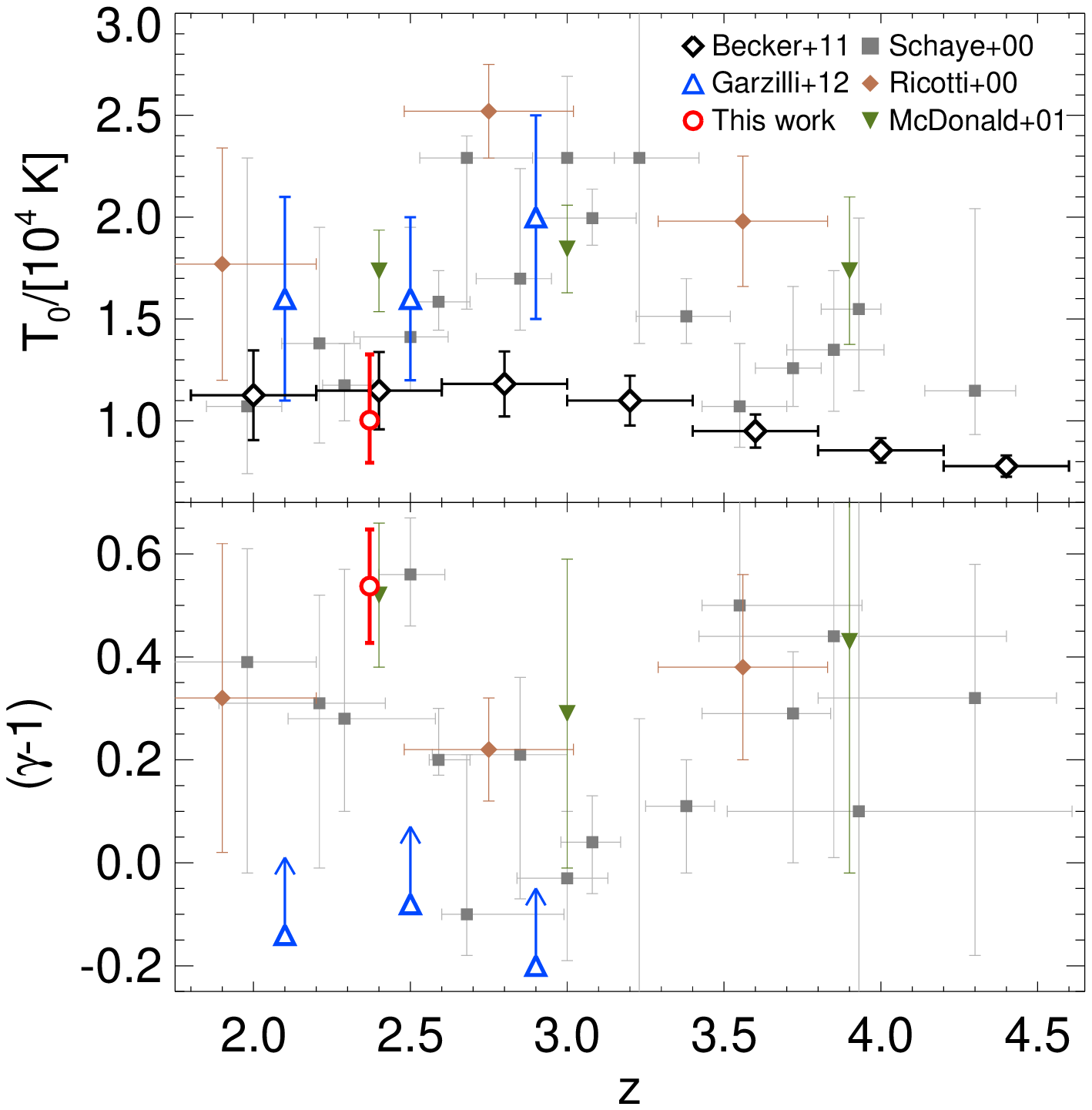}
  \vspace{-0.5cm}
  \caption{A comparison of the temperature at mean density, $T_{0}$
    (top panel), and slope of the $T$--$\Delta$ relation, $(\gamma-1)$
    (bottom panel), inferred in this work (red circles) to other
    recent constraints obtained using a variety of different methods
    at $2\leq z \leq 4.5$.  These are: the curvature statistic from
    \citet{Becker11} (black diamonds); the wavelet amplitude PDF from
    \citet{Garzilli12} (blue triangles), and the $b_{\rm HI}$--$N_{\rm
      HI}$ cut-off analyses presented by \citet{Schaye00} (grey
    squares), \citet{Ricotti00} (orange diamonds) and
    \citet{McDonald01} (inverted green triangles). All uncertainties
    are 1$\sigma$.  The $T_{0}$ values inferred from the measurements
    of \citet{Becker11} assume $(\gamma-1)=0.54\pm0.11$, i.e. the
    value inferred in this work at $z=2.4$.}
   
  \label{fig:temps}
\end{center}
\end{figure}

\begin{figure*}
  \centering
  \begin{minipage}{180mm}
    \begin{center}
      \psfig{figure=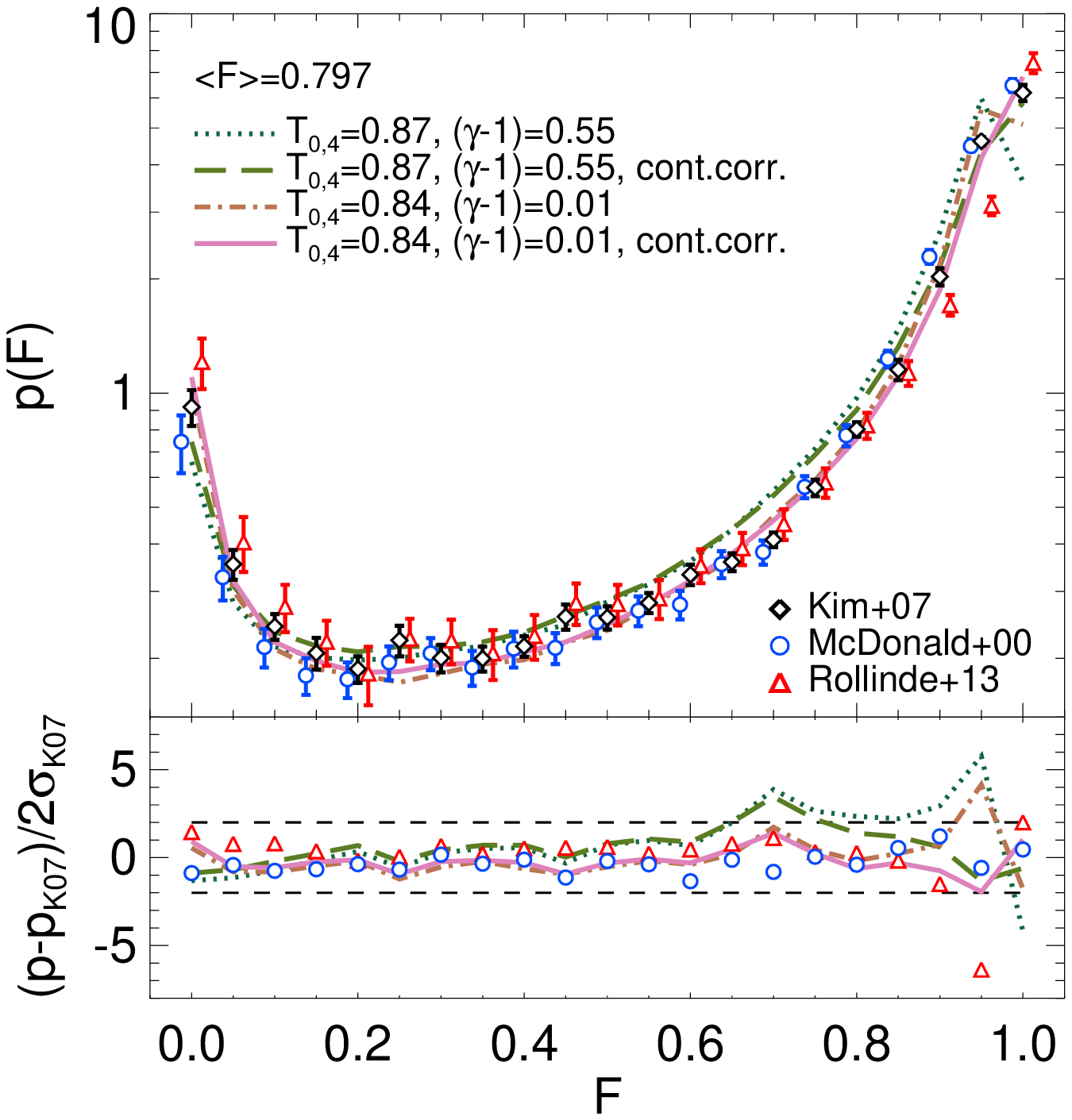,width=0.48\textwidth}
      \psfig{figure=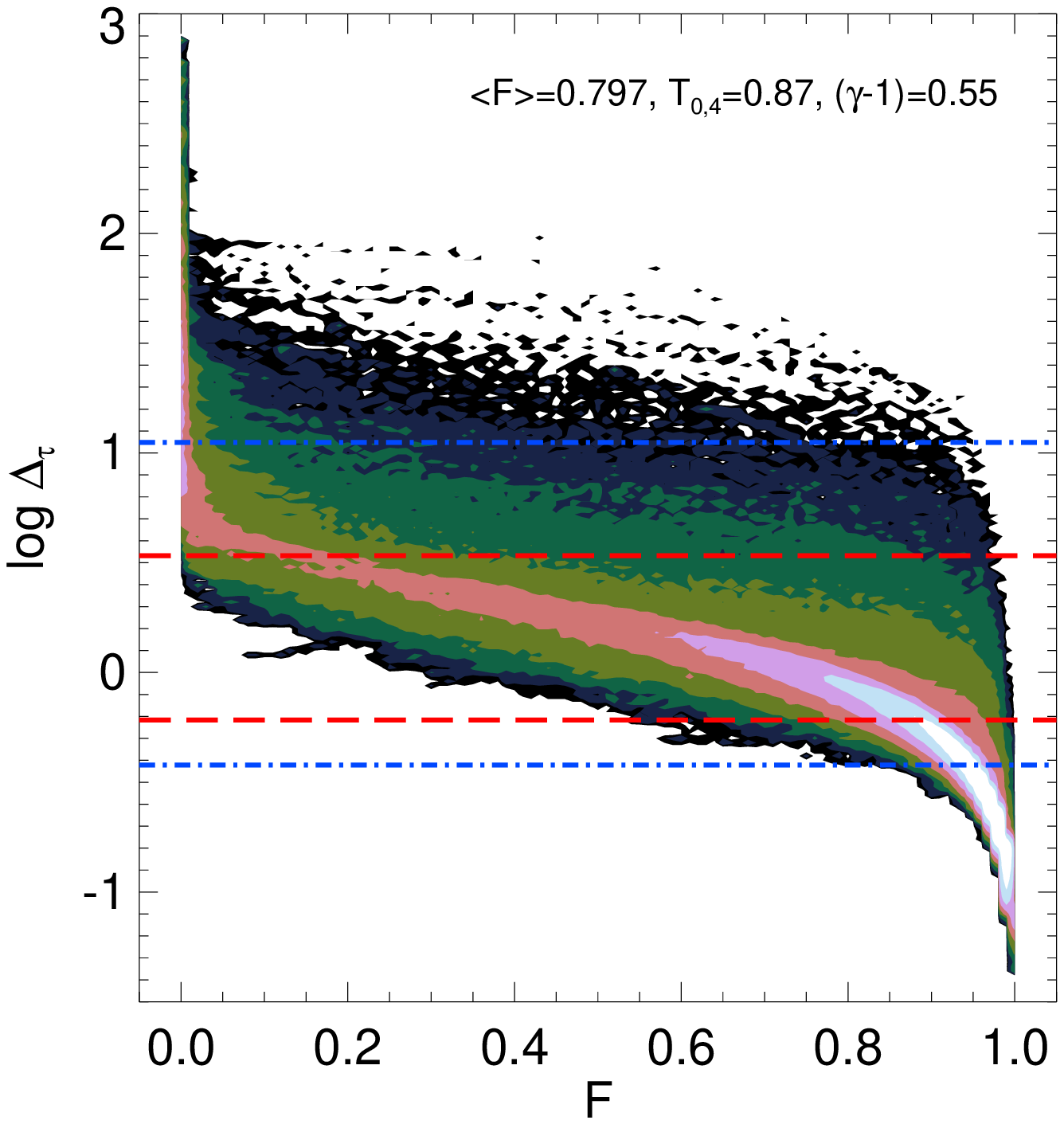,width=0.48\textwidth}

  \vspace{-0.2cm}
  \caption{{\it Top left:} The probability distribution of the
    transmitted flux (PDF) at $z\sim 2.5$.  The observational data are
    from K07 at $\langle z \rangle=2.52$, \citet{Rollinde13} at
    $\langle z \rangle=2.5$ and \citet{McDonald00} at $\langle z
    \rangle =2.4$.  The data points from the latter two studies have
    been offset by $\Delta F= \pm 0.0125$ for clarity of
    presentation. The curves display the PDF obtained from two
    different hydrodynamical simulations of the \Lya forest at
    $z=2.553$ which assume a $T$--$\Delta$ relation with either
    $(\gamma-1)=0.55$ or $(\gamma-1)=0.01$.  Each of these models is
    furthermore shown with and without an estimate for the continuum
    correction (see text for details). {\it Bottom left:} The
    difference between the mock data and observations from
    \citet{Rollinde13} and \citet{McDonald00} with respect to K07,
    normalised by twice the 1$\sigma$ K07 jack-knife errors.  This
    accounts for a possible factor of two underestimate in the sample
    variance suggested by \citet{Rollinde13}.  The horizontal dashed
    lines display the 2$\sigma$ bound for these increased error
    estimates.  {\it Right:} Contour plot of the optical depth
    weighted overdensity against the transmitted flux in each pixel of
    the mock spectra drawn from the simulation with $T_{0,4}=0.87$ and
    $(\gamma-1)=0.55$.  The number density of pixels increases by
    $0.5$ dex within each contour level.  The horizontal red dashed
    (blue dot-dashed) lines bound 68 (95) per cent of the
    overdensities corresponding to absorption lines with
    $10^{12.5}\leq N_{\rm HI}/[{\rm cm^{-2}}] \leq 10^{14.5}$.}
  \label{fig:PDF}

 \end{center}
  \end{minipage}
\end{figure*}

Our recalibrated measurements of the $T$--$\Delta$ relation at $z\sim
2.4$ are compared to previous measurements in Fig.~\ref{fig:temps}.
The $T_{0}$ and $(\gamma-1)$ constraints obtained in this study are in
excellent agreement with the recent, independent measurement presented
by \cite{Becker11} using the curvature statistic.  These authors
directly measure $T(\bar{\Delta})=[2.54\pm0.13]\times 10^{4}\rm\,K$
($2\sigma$) at a characteristic overdensity of $\bar{\Delta}=4.4$ at
$z=2.4$.  The $T_{0}$ and $(\gamma-1)$ results rederived here from the
RSP12 data translate to a value of
$T(\bar{\Delta})=[2.22^{+0.80}_{-0.59}]\times 10^{4}\rm\,K$
(1$\sigma$), which is fully consistent with the \cite{Becker11} value.
Alternatively, translating the \cite{Becker11} $T(\bar{\Delta})$
measurement to a temperature at the mean density assuming
$(\gamma-1)=0.54\pm 0.11$ yields $T_{0} = [1.15\pm0.19] \times
10^{4}\rm\,K$ ($1\sigma$).  For the value of $(\gamma-1)$ measured
here, therefore, both the \cite{Becker11} measurements and the present
results are consistent with a value of $T_{0}$ near $1 \times
10^{4}\rm\,K$ at $z = 2.4$.  Note that \cite{Becker11} also strongly
constrain $T_{0}$ to be relatively low at $z > 4$, where the
characteristic overdensity probed by the curvature approaches the mean
density.  For example, at $z=4.4$ they find $T_{0}$ to be in the range
0.70--0.94 $\times 10^{4}\rm\,K$ (2$\sigma$) for $(\gamma-1) = $
1.3--1.5.  This implies only a moderate boost to the IGM temperature
at mean density during \HeII reionisation, although the precise
evolution of $T_{0}$ over $2.4 < z < 4.4$ will depend on the evolution
of $(\gamma-1)$ over these redshifts.

Our measurement of $T_{0}$ is also in reasonable agreement with those
of \cite{Garzilli12}, who performed an analysis of the wavelet
amplitude PDF (see also \citealt{Lidz10}), obtaining
$T_{0}=[1.6\pm0.4]\times10^{4}\rm\,K$ ($1\sigma$) at the slightly
higher redshift of $z=2.5$.  We are also consistent with the earlier
$b_{\rm HI}$--$N_{\rm HI}$ measurements obtained by \cite{Schaye00} at
$z=2.4$; as already discussed these authors used a similar technique
but smaller data set compared to RSP12.  We are unable to directly
compare the $b_{\rm HI}$--$N_{\rm HI}$ measurements from
\cite{Ricotti00} to our measurement, since these authors do not
present results at $z=2.4$.  However, they find a signficantly larger
value of $T_{0}=25200\pm2300\rm\,K$ at $z=2.75$, which would require a
$T$--$\Delta$ relation slope of $(\gamma-1)\sim 1$ for consistency
with the \cite{Becker11} measurements at the same redshift.  Finally,
\cite{McDonald01}, again using the $b_{\rm HI}$--$N_{\rm HI}$ cut-off
(although with a different method to \citealt{Schaye00}), infer
$T=2.26\pm0.19\times10^{4}\rm\,K$ at $\Delta=1.66\pm0.11$.  Although
these authors do not present a measurement of $T_{0}$, using their
measurement of $(\gamma-1)=0.52\pm0.14$ this corresponds to
$T_{0}=1.74\pm0.20\times10^{4}\rm\,K$.  This temperature is somewhat
higher than the three other measurements at $z=2.4$, by
$2$--$3\sigma$.  Despite this, however, there appears to be a
reasonable consensus on the thermal state of the low-density IGM at
$z\sim 2.4$ from both \Lya transmission statistics \emph{and} line
decomposition analyses.

On the other hand, while our revised value of $(\gamma-1)=0.54\pm
0.11$ is fully consistent with the original line decomposition
analyses of \cite{Schaye00} and \cite{McDonald01} and the lower limits
from the wavelet PDF obtained by \cite{Garzilli12}, as noted by RSP12
there appears to be some tension with constraints based on the
observed transmission PDF (\citealt{Kim07}, hereafter K07) which
appear to favour an isothermal or perhaps even inverted $T$--$\Delta$
relation at $2<z<3$ (\citealt{Bolton08,Viel09}).  In the left hand
panel of Fig.~\ref{fig:PDF} we revisit this by comparing the PDF of
the transmitted flux from two of our hydrodynamical simulations to
three observational measurements of the \Lya forest transmission PDF
from high resolution ($R\sim40\,000$) data at $z\simeq2.4$--$2.5$
(K07, \citealt{McDonald00,Rollinde13}).  The mock spectra have been
drawn from models with a $T$--$\Delta$ relation with either
$(\gamma-1)=0.55$ or $(\gamma-1)=0.01$ at $z=2.553$, and have a
$T_{0}$ value close to the revised measurement obtained in this work.
The spectra have been processed following a similar procedure to that
outlined in Section~\ref{sec:sims}, with two key differences.
Firstly, the spectra are scaled to match the mean transmission of the
K07 PDF data, $\langle F \rangle =0.797$ at $\langle z \rangle=2.52$,
and Gaussian distributed noise matching the variances, $\sigma_{\rm
  F}$, in each bin of the K07 PDF is added.  Secondly, for each model
we also implement an iterative continuum correction to the spectra to
mimic the effect of a possible continuum placement error on the K07
data.  We first compute the median transmitted flux in each of our
mock sight-lines and then deselect all pixels below $1\sigma_{\rm F}$
of this value.  This procedure is then repeated for the remaining
pixels until convergence is achieved. The final median flux is
selected as the new continuum level.

The comparison in Fig.~\ref{fig:PDF} demonstrates a $T$--$\Delta$
relation with $(\gamma-1)=0.55$ is inconsistent with the K07 data at
$4$--$7\sigma$ at $0.6 <F < 0.8$, even when accounting for
possible continuum misplacement (although see the analysis presented
by \citealt{Lee12} using an analytical model for the transmission
PDF).  An isothermal model on the other hand is in within
$1$--$3\sigma$ of the K07 data over the same interval.  This is fully
consistent with the more detailed analysis performed by \cite{Viel09}
and \cite{Bolton08}.  Note, however, that \cite{Rollinde13} recently
argued that the error bars quoted on the K07 PDF measurements may be
larger by up to a factor of two simply due to sample variance; these
authors demonstrated the bootstrap or jack-knife errors on the data
will be underestimated if the chunk size which the spectra are sampled
over is too small.  Comparison of the three observational measurements
in Fig.~\ref{fig:PDF} likewise suggests that the errors on the PDF
measurements have been initially underestimated.

The lower panel in Fig.~\ref{fig:PDF} therefore displays the
difference between the K07 PDF and the mock data and observations from
\citet{Rollinde13} and \citet{McDonald00}, divided through by
\emph{twice} the 1$\sigma$ jack-knife errors reported by K07.  With
these larger uncertainties the discrepancy with the $(\gamma-1)=0.55$
model decreases, although the difference between the simulation and
data is still $\sim 2$--$3\sigma$ at $0.6<F<0.8$ after accounting for
a plausible continuum placement error.  This is in reasonable
agreement with \cite{Rollinde13}, who also reproduce the K07 PDF at
$z=2.5$ to within 2--3$\sigma$ of the expected
dispersion\footnote{This excludes the \cite{Rollinde13} $z=2.5$ PDF
  bin at $F=0.95$, which is discrepant with K07 at $\sim6\sigma$ even
  after doubling the K07 1$\sigma$ error bars.  \cite{Rollinde13}
  attribute large differences at $F>0.7$ to continuum uncertainties,
  although our analysis suggests that a reasonable estimate for the
  continuum misplacement may still not fully explain this offset at
  $F=0.95$.}  estimated from mock spectra drawn from the GIMIC
simulation suite (\citealt{Crain09}) assuming $(\gamma-1)\sim 0.35$
and $\langle F \rangle =0.770$.  \cite{Rollinde13} conclude there is
no evidence for a significant departure from a power-law $T$--$\Delta$
relation with $(\gamma-1)>0$.  The remaining disagreement between the
observations and a model with $(\gamma-1)=0.55$ suggests that
increased errors due to continuum placement and sample variance may
still not fully account for the true uncertainties on the
observational data.

In this context, it is worth stressing that the transmission PDF and
$b_{\rm HI}$--$N_{\rm HI}$ cut-off are sensitive to a different range
of IGM gas densities and hence also temperatures.  The right-hand
panel of Fig.~\ref{fig:PDF} displays the relationship between the
optical depth weighted overdensity and transmitted flux in each pixel
of the spectra drawn from the simulation with $(\gamma-1)=0.55$, shown
in the left panel.  For comparison, the red dashed (blue dot-dashed)
lines bound 68 (95) per cent of the overdensities probed corresponding
to absorption lines with $10^{12.5}\leq N_{\rm HI}/[{\rm cm^{-2}}]
\leq 10^{14.5}$ (i.e. the range used to measure the $b_{\rm
  HI}$--$N_{\rm HI}$ cut-off at $z\sim 2.4$).  Interestingly, the
largest differences between models and observations of the PDF arise
almost exclusively in the underdense regions which are not well
sampled by the $b_{\rm HI}$--$N_{\rm HI}$ cut-off measurements.
Furthermore, note that the $b_{\rm HI}$--$N_{\rm HI}$ cut-off is only
sensitive to the coldest gas lying along the lower bound of the
$T$--$\Delta$ plane (e.g. S99). It therefore remains possible that a
more complicated, multiple valued $T$--$\Delta$ plane with significant
scatter and/or a separate hot IGM component confined to the most
underdense regions, $\log \Delta_{\tau} \sim -0.5$, may also influence
the shape of the PDF at $F>0.7$.  Indeed, recent radiative transfer
simulations of \HeII reionisation predict significant scatter or even
bimodality in the $T$--$\Delta$ plane at $z\sim 3$
(\citealt{MeiksinTittley12,Compostella13}).  More detailed studies of
both the PDF and the $b_{\rm HI}$--$N_{\rm HI}$ distribution over a
wider redshift range, combined with simulations which have increased
dynamic range and/or incorporate radiative transfer effects, will
therefore be required to establish the importance of these effects.


\section{Conclusions} \label{sec:discuss}

We have performed a careful calibration of the measurement of the
$T$--$\Delta$ relation in the low-density IGM at $z \sim 2.4$.  Our
analysis is based on the $b_{\rm HI}$--$N_{\rm HI}$ cut-off measured
from mock \Lya forest spectra drawn from an extensive set of
high-resolution hydrodynamical simulations, combined with accurate
measurements of the mean transmitted flux in the \Lya forest
(\citealt{Becker13}) and the KBSS line profile fits from RSP12.  We
confirm the high value of the power-law slope, $(\gamma-1)$, at
$z\sim2.4$ advocated by RSP12, but we find a value for the temperature
at mean density, $T_0$, which is smaller by almost a factor of two.
The latter is mainly due to a difference of $0.65$ dex in the
calibration of the $N_{\rm HI}$--$\Delta$ correlation.  The lower
inferred value for the temperature brings the measurement of RSP12
into excellent agreement with the \cite{Becker11} constraint on the
IGM temperature at the same redshift, but inferred at somewhat higher
gas density using the curvature distribution of the transmitted flux.
More generally, recent IGM temperature measurements appear to now show
reasonable agreement and to favour the lower end of the range of
previously discussed values, suggesting that the heat injection into
the IGM during \HeII reionisation was moderate.

However, the high value of $(\gamma-1)$ which now appears to have been
measured with reasonable accuracy from the $b_{\rm HI}$--$N_{\rm HI}$
distribution at $z\sim 2.4$ disagrees with that inferred from the
transmitted flux PDF at $z \sim 2.5$ (at 2--3$\sigma$ for $0.6<F<0.8$)
even if assuming (i) previously reported uncertainties on the PDF have
been underestimated by a factor of two \emph{and} (ii) a plausible
estimate for the continuum placement uncertainty (see also
\citealt{Lee12,Rollinde13}). While it is possible this difference is
due to systematic uncertainties which remain underestimated, it is
important to emphasise that the $b_{\rm HI}$--$N_{\rm HI}$
distribution and PDF are sensitive to the temperature of the IGM in
nearly disjunct density ranges, with the PDF largely probing densities
below the mean.  While it appears very likely the IGM $T$--$\Delta$
relation is not inverted during \HeII reionisation from both an
observational and theoretical perspective
(e.g. \citealt{McQuinn09,Bolton09}), it may be too early to completely
discard the interesting possibility that our understanding of the
heating of the highly underdense IGM is still incomplete.

The convergence towards a self-consistent set of parameters describing
the $T$--$\Delta$ relation at $z \simeq 2.4$ represents an encouraging
step towards establishing the detailed thermal history of the
high-redshift IGM.  The extension of the measurements based on the
$b_{\rm HI}$--$N_{\rm HI}$ distribution to a wider redshift range, and
the development of methods that can push the temperature measurements
to lower density and search for the predicted scatter or even
bimodality in the $T$--$\Delta$ plane during \HeII reionisation
(e.g. \citealt{MeiksinTittley12,Compostella13}) will hopefully lead to
a fuller characterisation of the thermal state of the IGM and its
evolution, and thus more generally to the properties of ionising
sources in the early Universe.

\section*{Acknowledgments}

The hydrodynamical simulations used in this work were performed using
the Darwin Supercomputer of the University of Cambridge High
Performance Computing Service (http://www.hpc.cam.ac.uk/), provided by
Dell Inc. using Strategic Research Infrastructure Funding from the
Higher Education Funding Council for England.  We thank Volker
Springel for making GADGET-3 available, Bob Carswell for advice on
VPFIT and the anonymous referee for a report which helped improve this
paper.  The contour plots presented in this work use the cube helix
colour scheme introduced by \cite{Green11}. JSB acknowledges the
support of a Royal Society University Research Fellowship. GDB
acknowledges support from the Kavli Foundation. MGH acknowledges
support from the FP7 ERC Grant Emergence-320596.  MV is supported by
the FP7 ERC grant ``cosmoIGM'' and the INFN/PD51 grant.


\appendix
\section{Convergence test}

\begin{figure}
\begin{center}
  \includegraphics[width=0.48\textwidth]{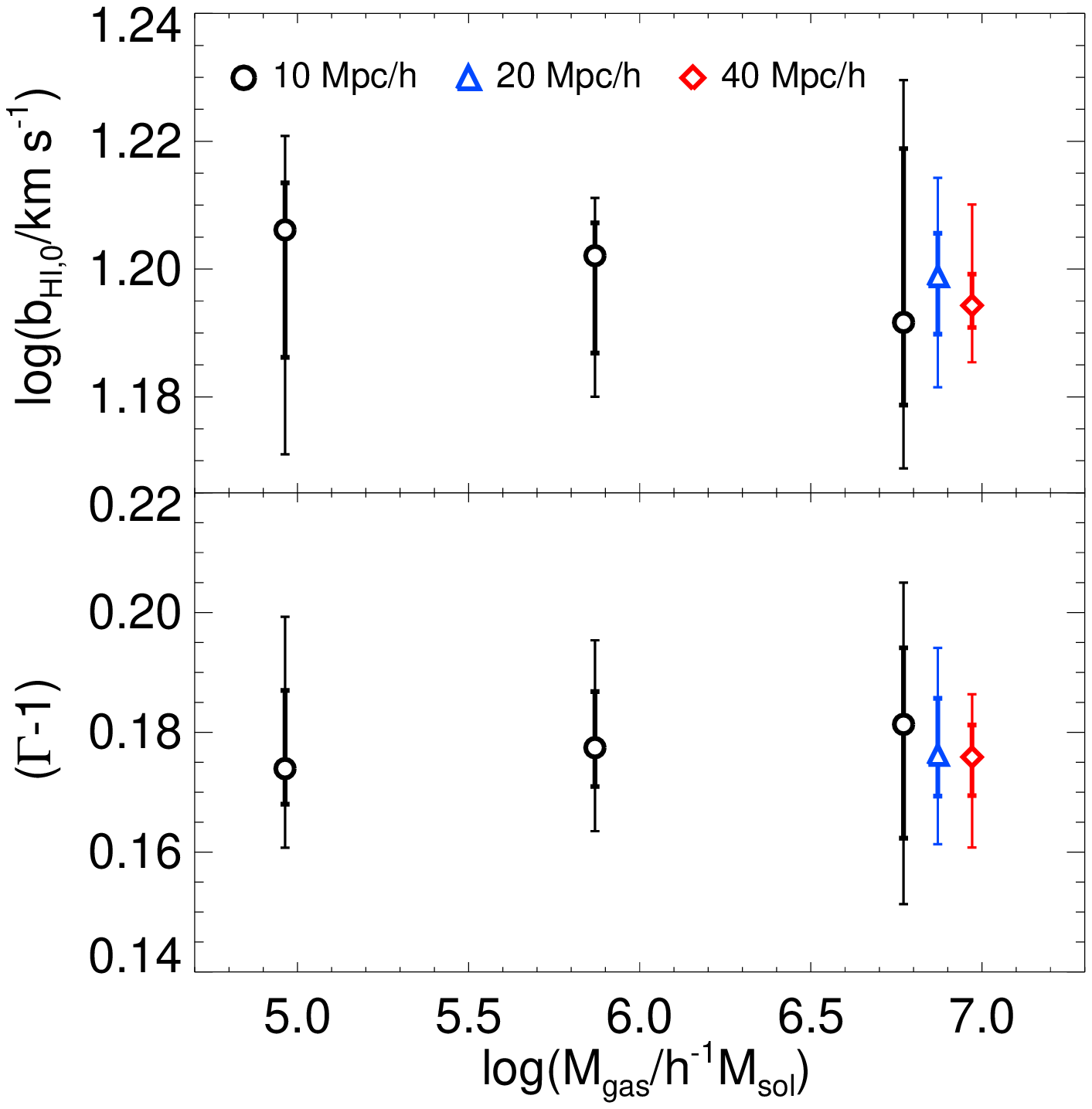}
  \vspace{-0.5cm}
  \caption{The amplitude, $b_{\rm HI,0}$ (top panel), and slope,
    $(\Gamma-1)$ (bottom panel) of the $b_{\rm HI}$--$N_{\rm HI}$
    cut-off measured from simulations with either a fixed box size
    ($10h^{-1}\rm\,Mpc$, black circles) or fixed mass resolution
    ($M_{\rm gas}=5.9\times 10^{6}h^{-1}M_{\odot}$) using a sample of
    $4\,000$ \Lya absorption lines.  For clarity of presentation the
    blue triangles and red diamonds have been offset in $\log M_{\rm
      gas}$ by $0.1$ and $0.2$ dex, respectively.  The thick (thin)
    bootstrap error bars correspond to the 68 (95) per cent confidence
    intervals around the median. The fiducial box size and mass
    resolution used in this work are $10h^{-1}\rm\,Mpc$ and $M_{\rm
      gas}=9.2\times 10^{4}h^{-1}M_{\odot}$.}

  \label{fig:converge}
\end{center}
\end{figure}

In Fig.~\ref{fig:converge} we present a convergence test of our
results by applying the $b_{\rm HI}$--$N_{\rm HI}$ cut-off algorithm
to five simulations with different box sizes and gas particle masses.
One of the simulations corresponds to the fiducial box size
($10h^{-1}\rm\,Mpc$) and gas particle mass resolution ($M_{\rm
  gas}=9.2\times 10^{4}h^{-1}M_{\odot}$) used in this work.  Two
further models test the mass resolution within a $10h^{-1}\rm\,Mpc$
box, assuming $M_{\rm gas}=7.4\times 10^{5}h^{-1}M_{\odot}$ and
$5.9\times 10^{6}h^{-1}M_{\odot}$, respectively. The final two models
test convergence with simulation volume, and have a fixed gas particle
mass of $M_{\rm gas}=5.9\times 10^{6}h^{-1}M_{\odot}$ within
$20h^{-1}\rm\,Mpc$ and $40h^{-1}\rm\,Mpc$ boxes.  These five
simulations are also described in \cite{Becker11}, and correspond to
models C15 and R1--R4 in their table 2.  

Mock spectra extracted at $z=2.355$ were analysed using the procedure
described in Section~\ref{sec:method} using a sample of $4\,000$
lines, with uncertainties on $b_{\rm HI,0}$ and $(\Gamma-1)$ estimated
by bootstrap sampling with replacement $4\,000$ times.  The
measurements of $b_{\rm HI,0}$ and $(\Gamma-1)$ are consistent within
the 68 per cent bootstrapped confidence intervals around the median,
indicating that our results should be well converged with box size and
mass resolution.  We have further verified that a smaller sample of
300 lines (e.g. S99) significantly increases the bootstrapped
uncertainties but does not introduce a systematic offset to the
results.

\end{document}